\documentclass[nohyper,12pt,letterpaper]{JHEP3}
\usepackage{epsfig}

\author{Robert de Mello Koch$^{1,2}$, Norman Ives$^1$, Jelena Smolic$^1$ and Milena Smolic$^1$\\
\qquad \\
Department of Physics and Centre for Theoretical Physics,$^1$\\ 
University of the Witwatersrand,\\ 
Wits, 2050,\\ 
South Africa\\
\qquad\\
Stellenbosch Institute for Advanced Studies,$^2$\\
Stellenbosch,\\
South Africa\\
\qquad\\
E-mail: \email{robert@neo.phys.wits.ac.za, ivesn@science.pg.wits.ac.za,  smolicj@science.pg.wits.ac.za, smolicm@science.pg.wits.ac.za}}

\abstract{
We find giant graviton solutions in Frolov's three parameter generalization of the Lunin-Maldacena background.
The background we study has $\tilde{\gamma}_1=0$ and $\tilde{\gamma}_2=\tilde{\gamma}_3=\tilde{\gamma}$. This
class of backgrounds provide a non-supersymmetric example of the gauge theory/gravity correspondence that can be
tested quantitatively, as recently shown by Frolov, Roiban and Tseytlin. The 
giant graviton solutions we find have a greater energy than the point gravitons, making 
them unstable states. Despite this, we find striking quantitative agreement between 
the gauge theory and gravity descriptions of open strings attached to the giant.}

\preprint{WITS-CTP-025}

\title{Unstable Giants}


\def \Tr{\mbox{Tr\,}}

\begin{document}

\section{Introduction}

The AdS/CFT correspondence\cite{AdSCFT} provides a new approach to the study of 
non-Abelian gauge theories.
One may hope that ultimately it may even be used to understand 
non-perturbative aspects of QCD, which is at the time of 
writing, a formidable problem. If this hope is ever to be realized, we must gain an 
understanding of the gauge theory/gravity correspondence in situations with no supersymmetry 
or conformal symmetry. Recently, a significant step in this direction was achieved by
Lunin and Maldacena\cite{LM}, who identified the gravitational dual of $\beta$ deformed 
${\cal N}=4$ super Yang-Mills theory. The dual gravitational theory has an AdS$_5$ times 
a deformed S$^5$ geometry. Since the AdS$_5$ factor is not deformed, the field theory is 
still conformally invariant. However, it only has ${\cal N}=1$ supersymmetry. This 
deformation was further generalized by Frolov\cite{F} who gave a background determined 
by three parameters, that in general, preserves no supersymmetry. The gauge theory/gravity
correspondence for this background was explored in detail by Frolov, Roiban and 
Tseytlin\cite{FRT2}. These authors went on
to show a quantitative agreement between the semi-classical energies of strings with
large angular momentum and the 1-loop anomalous dimensions of the corresponding gauge
theory operators. This is a significant result. The gauge
theory/gravity correspondence is a strong weak coupling duality in the 't Hooft coupling. At 
weak coupling computations in the field theory are straight forward; the dual gravitational 
theory however, has a highly curved geometry. At strong coupling computations in the field 
theory are not (in general) under control; in this case curvature corrections in the dual
gravitational theory can be neglected. The correspondence is usually explored by computing
``nearly protected quantities." These can be computed at weak coupling in the field theory. 
Since they are nearly protected, they can reliably be extrapolated to the strong coupling 
regime where comparison with the dual gravity theory is possible. Typically, one appeals to 
the supersymmetry of the problem to find these nearly 
protected quantities. The agreement of \cite{FRT2}
is striking because it provides an example of quantitative agreement between the gravity and 
field theory descriptions, in a setting without any supersymmetry. It is important to see 
how far this quantitative agreement in non-supersymmetric settings can be extended. This is 
the primary motivation for our work.

Giant gravitons\cite{Lenny},\cite{AdS},\cite{djm} provide a very natural 
framework for the study of non-perturbative effects
in the string theory, in supersymmetric examples of the gauge theory/gravity correspondence.
Since giant gravitons are BPS objects, they lead to effects that are protected and hence may 
be extrapolated between strong and weak coupling. Moreover, they have a simple description
in terms of a string worldsheet theory - to leading order they simply determine the boundary
conditions for strings with no other affect on the worldsheet sigma model. 
A lot is also known about giant gravitons in the dual field theory. Operators dual
to giant gravitons have been studied in both the $U(N)$\cite{CJR} and the $SU(N)$\cite{gwyn}
gauge theories. These half BPS states also have a simple description in terms of free fermions
for a one matrix model\cite{toy} which has recently been connected to a description which
accounts for the full back reaction of the geometry in the supergravity limit\cite{llm}.
A tantalizing attempt to go beyond one matrix dynamics has appeared in\cite{joao}.
Further, the technology needed to study strings attached to giant gravitons
is well developed\cite{bb},\cite{bv}. Given the recent
progress in constructing non-supersymmetric examples of the gauge theory/gravity correspondence,
it seems natural to ask if there are giant gravitons solutions in these new geometries. We
will construct giant gravitons for the deformed
background with $\tilde{\gamma}_1=0$ and $\tilde{\gamma}_2=\tilde{\gamma}_3=\tilde{\gamma}$.

A particularly efficient way to organize and sum the Feynman diagrams of the field theory, 
is through the use of a spin chain\cite{MZ}. In this approach, one identifies the 
dilatation operator 
of the field theory with the Hamiltonian of the spin chain. Constructing operators with a 
definite scaling dimension as well as the spectrum of scaling dimensions becomes the problem
of diagonalizing the spin chain Hamiltonian. This approach has been extremely powerful because 
it allows one to identify and match the integrability of the gauge theory 
dilatation operator\cite{BKM}
with that of the world sheet sigma model\cite{MSW}.
Understanding the field theory beyond the one-loop approximation involves studying spin chains
with varying number of sites\cite{Beisert}. In this article 
we would like to use the spin chain approach to 
study operators dual to open strings attached to giant gravitons. For non-maximal giants this 
again corresponds to studying a spin chain with a variable number of sites. A very convenient
approach to these problems has been developed in\cite{BCV}. The 
idea is to map the spin chain into a 
dual boson model on a lattice. For the boson model, the number of sites is fixed; the variable
number of sites in the original spin chain is reflected in the fact that the number of bosons 
in the dual boson model is not conserved. In this article we will construct the boson lattice
model which describes open strings attached to giant gravitons in the deformed background.

Apart from the three parameter deformed backgrounds studied in this article, there have been many other
interesting developments following Lunin and Maldacena's work. In\cite{FRT}
energies of semiclassical string states in the Lunin Maldacena background 
were matched to the anomalous dimensions of a class of gauge theory scalar operators. 
The spin chain for the twisted ${\cal N}=4$ super Yang-Mills has been studied in \cite{BR}. 
The logic employed by Lunin and Maldacena to obtain the gravitational theory dual to the
deformed field theory has been extended in a number of ways. Recently, instead of deforming
the ${\cal N}=4$ super Yang-Mills theory, deformations of ${\cal N}=1$ and ${\cal N}=2$
theories have been considered\cite{GN}. Further, deformations of eleven
dimensional geometries of the form AdS$_4\times$Y$_7$ with $Y_7$ a seven dimensional
Sasaki-Einstein\cite{CA},\cite{Lee} or weak $G_2$ or tri-Sasakian\cite{Lee} 
space have been considered. The pp-wave limit of the Lunin Maldacena background,
and the relation to BMN\cite{BMN} operators in the dual field theory has been considered in \cite{NP}. 
Recent studies of the $\beta$-deformed field theory include\cite{ft}. Semiclassical
strings were studied in \cite{last}.
Finally, in \cite{Russo}, interesting instabilities in the general three parameter
backgrounds have been discovered.

Our paper is organized as follows:
In the next section we give an ansatz for the giant graviton solutions. These giant gravitons
blow up in the deformed $S^5$ of the geometry. We compute the energy and show that the energy
of the point graviton is lower than that of the giant graviton, making the giant graviton
an unstable state. In section 3 we explicitly demonstrate that the giant graviton extremizes
the action. Further, we study vibration modes of the giant arising from the excitation of the
AdS$_5$ coordinates. In contrast to the AdS$_5\times$S$^5$ vibration spectrum, we find that
the frequencies of these modes does depend on the radius of the giant. We recover the 
AdS$_5\times$S$^5$ vibration spectrum for large giants. Our results show that the giant graviton
is perturbatively stable. We construct a bounce solution to the Euclidean equations of motion,
demonstrating that the giant graviton is corrected by quantum tunneling. In section 4 we compute the 
Hamiltonian of the lattice boson model. The energies of this Hamiltonian give the anomalous dimensions
of the operators dual to open strings ending on the giant. Using coherent states we obtain an
action governing the semiclassical dynamics of these strings. We find complete agreement with
the semiclassical dynamics following from the dual string sigma model. In section 5 we 
summarize and discuss our results.

\section{Giant Graviton Solutions}

In this section we will obtain giant graviton solutions in the deformed background.
The giant graviton solutions we consider are D3 branes that have blown up in the deformed sphere
part of the geometry. Our ansatz for the giant, made at the level of the action, assumes that it 
has a constant radius and a constant angular velocity. This ansatz will be justified in section 3 
where we will argue that our solution does indeed extremize the action.

To write down the action for the D3 brane, we need the metric and dilaton of the background (to write down
the Dirac-Born-Infeld term in the action), the NS-NS two form potential and the RR two and four form 
potentials (to write down the Chern-Simons terms
in the action). The $AdS_5$ and the deformed sphere spaces are orthogonal to each other

$$ ds^2 =ds^2_{AdS_5}+ds^2_{S^5_{def}}.$$

\noindent
We will use the following spacetime coordinates:

\noindent
(1) For $AdS_5$ use $(t,\alpha_1,\alpha_2,\alpha_3,\alpha_4)$. In terms of these coordinates, the metric is

$$ ds^2_{AdS_5}=-\left(1+\sum_{k=1}^4\alpha_k^2\right)dt^2+R^2\left(\delta_{ij}
+{\alpha_i\alpha_j \over 1+\sum_{k=1}^4\alpha_k^2}\right)d\alpha_i d\alpha_j .$$

\noindent
These coordinates are useful when studying small fluctuations of the giant graviton, since the make the $SO(4)$
subgroup of the $SO(2,4)$ isometry of $AdS_5$ manifest.

\noindent
(2) For the deformed five sphere, use $(\alpha,\theta,\phi_1,\phi_2,\phi_3 )$. In terms of these coordinates, the metric is

$$ds^2_{S^5_{def}}=R^2\left(
d\alpha^2 +\sin^2\alpha d\theta^2 
+G\sum_{i=1}^3\rho_i^2 d\phi_i^2\right)+R^2 G\rho_1^2\rho_2^2\rho_3^2
\left(\sum_{i=1}^3\tilde{\gamma}_i d\phi_i\right)^2,$$

$$\rho_1=\cos\alpha,\qquad \rho_2=\sin\alpha\cos\theta,\qquad \rho_3=\sin\alpha\sin\theta ,$$

$$ G^{-1}=1+\tilde{\gamma}_1^2\rho_2^2\rho_3^2+\tilde{\gamma}_2^2\rho_1^2\rho_3^2+\tilde{\gamma}_3^2\rho_2^2\rho_1^2 .$$

\noindent
In terms of the dilaton $\phi_0$ of the undeformed background, the dilaton is

$$e^\phi =\sqrt{G}e^{\phi_0}.$$

\noindent
The dilaton of the undeformed background satisfies $ R^4 e^{-\phi_0}=4\pi N l_s^4.$
The five form field strength of the background is

$$ F_5= 4R^4 e^{-\phi_0}\left(\omega_{AdS_5}+G\omega_{S^5}\right),\qquad
\omega_{S^5}=\cos\alpha\sin^3\alpha\sin\theta\cos\theta d\alpha d\theta 
d\phi_1 d\phi_2 d\phi_3 .$$

\noindent
Finally, the RR two form potential is

$$ C_2=-4R^2 e^{-\phi_0}\omega_1 d\left(\sum_{i=1}^3\tilde{\gamma}_i\phi_i\right),\qquad
d\omega_1=\cos\alpha\sin^3\alpha\sin\theta\cos\theta d\alpha d\theta ,$$

\noindent
and NS-NS two form potential is

$$B=R^2 G\omega_2,\qquad
\omega_2=\tilde{\gamma}_3\rho_1^2\rho_2^2d\phi_1d\phi_2+
\tilde{\gamma}_1\rho_2^2\rho_3^2d\phi_2d\phi_3+
\tilde{\gamma}_2\rho_3^2\rho_1^2d\phi_3d\phi_1 .$$

\noindent
We will not consider the most general background with three arbitrary parameters in this paper;
from now on we set $\tilde{\gamma}_1=0$ and $\tilde{\gamma}=\tilde{\gamma}_2=\tilde{\gamma}_3$.

To write down the D3 brane action 

$$S=-{1\over (2\pi )^3 l_s^4}\int d^4 y\, e^{-\phi}\sqrt{|\det(G+B)|} +\int C_4 +\int C_2\wedge B ,$$

\noindent
we will use static gauge

$$y^0=t,\quad y^1=\theta,\quad y^2=\phi_2,\quad y^3=\phi_3 .$$

\noindent
Our ansatz for the giant graviton is $\alpha=\alpha_0$, $\phi_1 =\omega t$ where
$\alpha_0$ and $\omega$ are constants, independent of $y^\mu $. It is now a simple matter
to integrate the Lagrangian density over $y^1$, $y^2$ and $y^3$ to obtain the Lagrangian 

$$L=-m\sqrt{1-a\dot{\phi}^2_1}+b\dot{\phi}_1,$$

\noindent
where

$$m= 2\pi^2 r^3 {e^{-\phi_0}\over (2\pi)^3 l_s^4} = N{r^3\over R^4} ,\qquad
a=R^2-r^2,$$

\begin{eqnarray}
b&=&4N\left[
{\tilde{\gamma}-\sqrt{4+\tilde{\gamma}^2}\over 4\tilde{\gamma}^2\sqrt{4+\tilde{\gamma}^2}}
\log\left(
{{2{r^2\over R^2}+{\sqrt{4+\tilde{\gamma}^2}\over\tilde{\gamma}}-1}\over
{{\sqrt{4+\tilde{\gamma}^2}\over\tilde{\gamma}}-1}}\right)-
{{\tilde{\gamma}+\sqrt{4+\tilde{\gamma}^2}\over 4\tilde{\gamma}^2\sqrt{4+\tilde{\gamma}^2}}}
\log\left(
{{-2{r^2\over R^2}+{\sqrt{4+\tilde{\gamma}^2}\over\tilde{\gamma}}+1}\over
{{\sqrt{4+\tilde{\gamma}^2}\over\tilde{\gamma}}+1}}\right)\right]
\nonumber\cr
&-&N{\tilde{\gamma}^2{r ^6 (R^2 -r^2)\over R^8 (1+\tilde{\gamma}^2{r^2\over R^2}(1-{r^2\over R^2}))}.}
\nonumber
\end{eqnarray}

\noindent
$r\equiv R\sin\alpha_0$ is the radius of the giant, 
$T_3$ is the D3 brane tension and $R$ is the radius of curvature of
the AdS space and the radius of the (undeformed) sphere.
As a check of our normalizations, we have verified that we recover the undeformed Lagrangian\cite{Lenny}
for giant gravitons in AdS$_5\times$S$^5$ in the $\tilde{\gamma}\to 0$ limit. Solving for $\dot{\phi}_1$
in terms of the angular momentum

$${\cal M}={\partial L\over\partial\dot{\phi}_1},$$

\noindent
we obtain

\begin{equation}
\dot{\phi}_1=\pm {{\cal M}-b\over\sqrt{a\big[{\cal M}-b\big]^2+m^2 a^2}}.
\label{omega}
\end{equation}

\noindent
The energy of the giant graviton is now easily computed

$$E=\dot{\phi}_1{\cal M}-L=\sqrt{m^2 + {\big[{\cal M}-b\big]^2\over a}}.$$

\noindent
We determine $\alpha_0$ by minimizing the energy at fixed ${\cal M}$. 

{\vskip -0.6cm}
\begin{center}
\begin{figure}[h]{\psfig{file=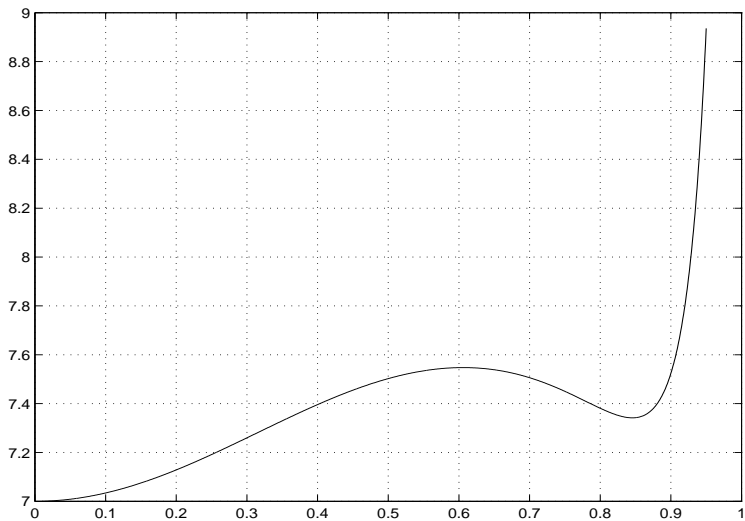,width=10cm,height=7cm}
 \caption{The energy of the giant graviton versus $r/R$ for fixed angular momentum. For the plot shown,
$\tilde{\gamma}=0.4$, $N=10$ and ${\cal M}=7/N$. The energy is shown in units of $1/R$.}}
\end{figure}
\end{center} 
{\vskip -0.6cm}

Clearly the energy of the point graviton is less than that of the giant, so that the giant graviton will be an
unstable state. We will study the nature of this instability in the next section. The contributions to the 
Chern-Simons four form flux and $C_2 \wedge B$ terms enter with opposite signs. At $\tilde{\gamma}=0$, the
$C_2 \wedge B$ term vanishes, while the four form flux term is non-zero. As $\tilde{\gamma}$ is increased,
the $C_2 \wedge B$ term grows faster than the four form flux term. For large enough deformations, the 
$C_2 \wedge B$ term dominates. There is a critical deformation beyond which there is no giant graviton solution. 
This matches well with the study \cite{shanin} of giants in a constant NSNS B field, in the maximally supersymmetric
type IIB-plane wave background. Other work on non-spherical giants and giants in a $B$ field include\cite{shanin2}.

{\vskip -0.6cm}
\begin{center}
\begin{figure}[h]{\psfig{file=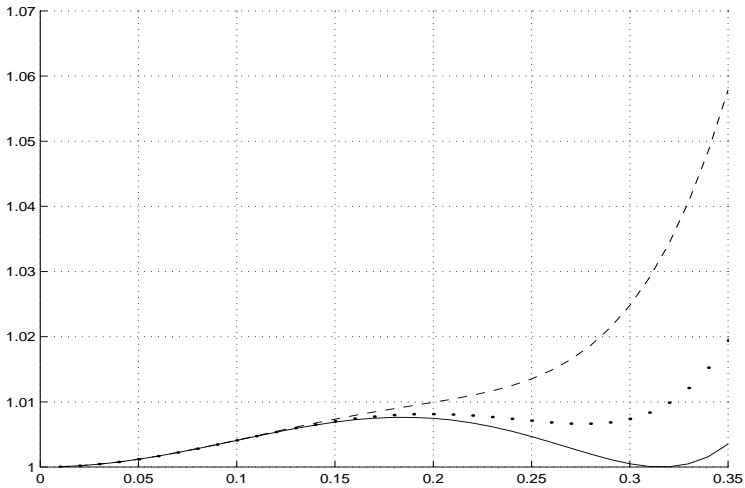,width=10cm,height=7cm}
 \caption{The energy of the giant graviton versus $r/R$ for fixed angular momentum. For the plot shown,
$N=10$ and ${\cal M}=1/N$. The solid line has $\tilde{\gamma}=0$, the dotted line $\tilde{\gamma}=0.8$ and
the dashed line $\tilde{\gamma}=1.6$. The energy is shown in units of $1/R$. As the deformation increases
the giant graviton minimum is raised until it is no longer a solution.}}
\end{figure}
\end{center} 
{\vskip -0.6cm}

\section{Fluctuations}

We have no guarantee that our ansatz of the previous section in fact minimizes the action. In this section
we check that this is indeed the case and further, we study the spectrum of certain vibration modes of the giant.
There are a number of interesting questions that can be answered using the vibration spectrum of giant gravitons.
If our giants belong to a family of solutions that all have the same energy and angular momentum, there will be modes
with zero energy. Secondly, if our giant graviton solution is (perturbatively) unstable, there will be tachyonic 
vibration mode(s). The excitations we consider correspond to motions of the branes in spacetime. Consequently, we
do not consider the possibility of exciting fermionic modes or gauge fields that live on the giant graviton's
worldvolume. Our results show that the giant graviton is perturbatively stable. Finally, we argue that the giant 
graviton is corrected by quantum tunneling by constructing a bounce solution to the Euclidean equations of motion.

Our ansatz for the giant graviton is

$$\alpha_i=\epsilon\delta\alpha_i,\qquad i=1,2,3,4,$$

$$\alpha =\alpha_0 +\epsilon\delta\alpha ,$$

$$\phi_1 =\omega t +\epsilon\delta\phi_1 .$$

\noindent
$\alpha_0$ and $\omega$ are constants, independent of $y^\mu $. Despite their names, we have not yet
given any reason to identify $\alpha_0$ and $\omega$ with the constants appearing in our ansatz of section 2. 
We now plug this ansatz into the action and expand in $\epsilon$.
If the linear order in $\epsilon$ contribution to the action vanishes, for $\omega =\dot{\phi}_1$ computed using
(\ref{omega}) and for the value of $\alpha_0$ that minimizes the energy, we know that the giants of section 2 
minimize the action and that they are indeed classical
solutions. The quadratic in $\epsilon$ contribution to the action can be used to learn about the energies
of vibration modes of the giant.

Plugging this ansatz into the action and expanding, the term linear in $\epsilon$ is

$$\epsilon\int dy^0 dy^1 dy^2 dy^3\left( A{\partial \delta\phi_1\over\partial t} +B\delta \alpha \right),$$

\noindent
where

\begin{eqnarray}
A&=& -{N\over 2\pi^2}\sin^3\alpha_0\sin y^1\cos y^1{\omega\cos^2\alpha_0\over\sqrt{1-\omega^2\cos^2\alpha}}
-N\sin y^1\cos y^1{\tilde{\gamma}^2{r ^6 (R^2 -r^2)\over 2\pi^2 R^8 (1+\tilde{\gamma}^2{r^2\over R^2}(1-{r^2\over R^2}))}}
\nonumber\cr
&+&{2N\sin y^1\cos y^1\over\pi^2}
\left[
{\tilde{\gamma}-\sqrt{4+\tilde{\gamma}^2}\over 4\tilde{\gamma}^2\sqrt{4+\tilde{\gamma}^2}}
\log\left(
{{2{r^2\over R^2}+{\sqrt{4+\tilde{\gamma}^2}\over\tilde{\gamma}}-1}\over
{{\sqrt{4+\tilde{\gamma}^2}\over\tilde{\gamma}}-1}}\right)
\right.\nonumber\cr 
&-&\left. {{\tilde{\gamma}+\sqrt{4+\tilde{\gamma}^2}\over 4\tilde{\gamma}^2\sqrt{4+\tilde{\gamma}^2}}}
\log\left(
{{-2{r^2\over R^2}+{\sqrt{4+\tilde{\gamma}^2}\over\tilde{\gamma}}+1}\over
{{\sqrt{4+\tilde{\gamma}^2}\over\tilde{\gamma}}+1}}\right)\right],
\nonumber
\end{eqnarray}

\begin{eqnarray}
B&=&
{N\cos \alpha_0\sin^2\alpha_0\sin y^1\cos y^1\over 4\pi^2 \sqrt{1-\omega^2\cos^2\alpha_0}}
(6\omega^2\cos^2\alpha_0-2\omega^2\sin^2\alpha_0-6)
+{2N\omega\cos\alpha_0\sin^3\alpha_0\sin y^1\cos y^1\over \pi^2(1+\tilde{\gamma}^2\cos^2\alpha_0\sin^2\alpha_0)}
\nonumber\\
&-&{N\omega\tilde{\gamma}^2(3\sin^5\alpha_0\cos^3\alpha_0-\sin^7\alpha_0\cos\alpha_0+2\tilde{\gamma}^2
\sin^7\alpha_0\cos^5\alpha_0)\sin y^1\cos y^1\over \pi^2(1+\tilde{\gamma}^2\cos^2\alpha_0\sin^2\alpha_0)^2}.
\nonumber
\end{eqnarray}

\noindent
Now, notice that the coefficient $A$ is independent of time. This implies that the term in the first order change in the 
action involving $\delta\phi_1$ gives no contribution, because we vary with fixed boundary conditions, that is, $\delta\phi_1$
vanishes at the initial and final times. Using (\ref{omega}) and plotting $B$ as a function of $\alpha_0$, we find that the
value of $\alpha_0$ that minimizes the energy is the same value of $\alpha_0$ that sets $B$ to zero. 

{\vskip -0.6cm}
\begin{center}
\begin{figure}[h]{\psfig{file=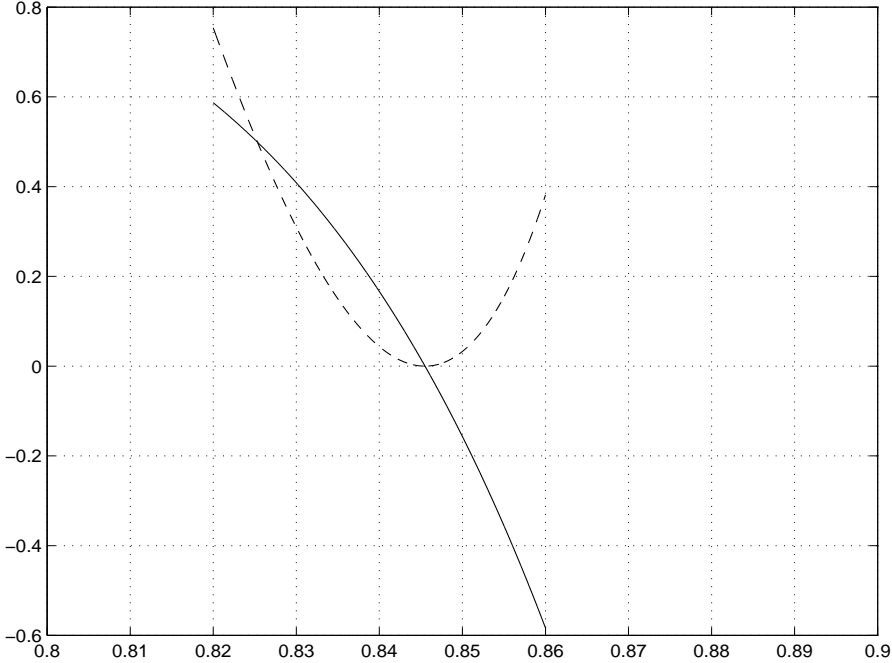,width=12cm,height=9cm}
 \caption{In the above plot $B$ is shown as the solid line; the energy of the giant graviton 
minus the minimum of the energy is shown as the dashed line. The $x$-axis is $r/R$. For the plot shown,
$\tilde{\gamma}=0.4$, $N=10$ and ${\cal M}=7/N$. The energy is shown in units of $1/R$.}}
\end{figure}
\end{center} 
{\vskip -0.6cm}

\noindent
This confirms that the giant gravitons written down in section 2 are indeed solutions to the equations of motion 
following from the D3 brane action. 

Expanding the action to second order in $\epsilon$ and varying with respect to $\delta\alpha_i$ we obtain the wave equation

$$ \partial^2_0\delta\alpha_i
+{1-\omega^2R^2 \cos^2\alpha_0\over R^2\sin^2\alpha_0}L^2\delta\alpha_i
-{\tilde{\gamma}^2\cos^2 (\alpha_0)\over R^2}(\partial_2-\partial_3)^2\delta\alpha_i
+{\delta\alpha_i\over R^2}=0,$$

\noindent
where we have introduced the angular momentum squared $L^2$, which in our coordinates is given by

$$-{2\over \sin (2y^1)}\left({1\over 2}\sin (2y^1)
{\partial\over\partial y^1}{\partial\over\partial y^1}
+\cos (2y^1) {\partial\over\partial y^1}
+\tan y^1{\partial\over\partial y^2}{\partial\over\partial y^2}
+\cot y^1{\partial\over\partial y^3}{\partial\over\partial y^3}\right).$$

\noindent
The original $SO(4)$ worldvolume symmetry that we'd have in the undeformed case is broken to $U(1)\times U(1)$.
These two $U(1)$ symmetries correspond to translations of $\phi_2$ and $\phi_3$. It is possible to choose spherical 
harmonics $Y^l_{m_1,m_2}(y^1,y^2,y^3)$
with definite $U(1)\times U(1)$ quantum numbers $(m_1,m_2)$. For spherical harmonics with $L^2=l(l+2)$ we 
have $|m_1|+|m_2|\le l$. Making the ansatz

$$\delta\alpha_i=e^{iE^l_{m_1,m_2} y^0}Y^l_{m_1,m_2}(y^1,y^2,y^3),$$

\noindent
we find 

$$(E^l_{m_1,m_2})^2 ={1\over R^2}+l(l+2)\left[{1-\omega^2 R^2 \cos^2\alpha_0\over R^2\sin^2\alpha_0}\right]
+{\tilde{\gamma}^2\cos^2 (\alpha_0)\over R^2}(m_1-m_2)^2. $$

\noindent
Clearly these frequencies depend on $R\sin\alpha_0$, the radius of the giant. For a near maximal giant, we
have $\sin\alpha_0\approx 1$ and $\cos\alpha_0\approx 0$, so that

$$(E^l_{m_1,m_2})^2 ={1\over R^2}+{l(l+2)\over R^2}. $$

\noindent
This is equal to the frequency obtained in \cite{djm} 
for giant gravitons in the undeformed AdS$_5\times$S$^5$ background.
Note that this frequency is independent of the size of the graviton. This is true for 
all giant gravitons (not just the maximal giant) in the undeformed background\cite{djm}.

Varying with respect to $\delta\phi_1$ and $\delta\alpha$ we obtain the following two (coupled) wave equations

$$ \partial^2_0\delta\alpha
+{1-\omega^2 R^2 \cos^2\alpha_0\over R^2\sin^2\alpha_0}L^2\delta\alpha
-{\tilde{\gamma}^2\cos^2 (\alpha_0)\over R^2}(\partial_2-\partial_3)^2\delta\alpha
+A_1\delta\alpha +A_2\partial_0\delta\phi_1=0,$$

$$ \partial^2_0\delta\phi_1
+{1\over R^2\sin^2\alpha_0}L^2\delta\phi_1
-{A_2\over\cos^2\alpha_0}\partial_0\delta\alpha=0,$$

\noindent
where

\begin{eqnarray}
A_1=-{2
\left(6\omega^2\cos^2\alpha_0\cot^2\alpha_0-6\cot^2\alpha_0-10\omega^2\cos^2\alpha_0+3+\omega^2\sin^2\alpha_0
+{\omega^4\sin^2\alpha_0\cos^2\alpha_0\over 1-\omega^2\cos^2\alpha_0}\right)\over\sqrt{1-\omega^2\cos^2\alpha_0}}
\nonumber\cr
+{4\omega\tilde{\gamma}^2 (15\sin\alpha_0\cos^4 \alpha_0-16\sin^3 \alpha_0\cos^2 \alpha_0
+\sin^5\alpha_0 +14\tilde{\gamma}^2\sin^3\alpha_0\cos^6\alpha_0
-10\tilde{\gamma}^2\sin^5\alpha_0\cos^4\alpha_0)
\over (1+\tilde{\gamma}^2\cos^2 (\alpha_0)\sin^2 (\alpha_0))^2}
\nonumber\cr
-{8\omega\tilde{\gamma}^4 (3\sin^3\alpha_0\cos^4 \alpha_0-\sin^5 \alpha_0\cos^2 \alpha_0
+2\tilde{\gamma}^2\sin^5\alpha_0\cos^6\alpha_0
(\cos^2\alpha_0 -\sin^2\alpha_0)
\over (1+\tilde{\gamma}^2\cos^2 (\alpha_0)\sin^2 (\alpha_0))^3}
\nonumber\cr
-8\omega\left(
{3\cos \alpha_0\cot\alpha_0 -\sin\alpha_0
\over 1+\tilde{\gamma}^2\cos^2 (\alpha_0)\sin^2 (\alpha_0)}
+{2\tilde{\gamma}^2 \cos^2 \alpha_0\sin\alpha_0(\sin^2\alpha_0-\cos^2\alpha_0)
\over (1+\tilde{\gamma}^2\cos^2 (\alpha_0)\sin^2 (\alpha_0))^2}
\right),
\nonumber
\end{eqnarray}

\begin{eqnarray}
2A_2&=&{4\omega\cos\alpha_0-10\omega\cos^3\alpha_0\over\sin\alpha_0\sqrt{1-\omega^2\cos^2\alpha_0}}+
{2\omega^3\cos^3\alpha_0\sin\alpha_0\over (1-\omega^2\cos^2\alpha_0)^{3/2}}
+4\omega\tilde{\gamma}^2{\sin^2\alpha_0\cos\alpha_0 (\sin^2\alpha_0-\cos^2\alpha_0)\over
(1+\tilde{\gamma}^2\cos^2 (\alpha_0)\sin^2 (\alpha_0))^2} \nonumber\cr
&-&{8\omega\tilde{\gamma}^2\cos^3\alpha_0\sin^2\alpha_0\over 
1+\tilde{\gamma}^2\cos^2 \alpha_0\sin^2 \alpha_0}
-{8\cos\alpha_0\over
1+\tilde{\gamma}^2\cos^2 \alpha_0\sin^2 \alpha_0}.
\nonumber
\end{eqnarray}

\noindent
When $\tilde{\gamma}=0$, $\omega =1$ and

$$A_1=0,\qquad A_2=-{2\cos\alpha_0\over\sin^2\alpha_0}.$$

\noindent
Using these values, it is easy to verify that we reproduce the undeformed results of \cite{djm}. Using the ansatz

$$\delta\alpha=A_\alpha e^{iE^l_{m_1,m_2} y^0}Y^l_{m_1,m_2}(y^1,y^2,y^3),\qquad
\delta\phi_1=A_\phi e^{iE^l_{m_1,m_2} y^0}Y^l_{m_1,m_2}(y^1,y^2,y^3),$$

\noindent
the energies of our fluctuations are found by solving

\begin{eqnarray}
{(E^l_{m_1,m_2})^4\over \sin^2\alpha_0}+{l(l+2)\over R^2\sin^2\alpha_0}\left({A_1\over R^2}+\tilde{\gamma}^2(m-n)^2{\cos^2\alpha_0\over R^2}+
{(1-\omega^2 R^2 \cos^2\alpha_0)\over R^2\sin^2\alpha_0}l(l+2)\right) \nonumber\cr
-(E^l_{m_1,m_2})^2
\left({\tilde{\gamma}^2(m-n)^2\cos^2\alpha_0+A_1+l(l+2)(2-\omega^2 R^2 \cos^2\alpha_0)\over R^2\sin^2\alpha_0} 
+{A_2^2(1-\omega^2 R^2 \cos^2\alpha_0)\over R^2\cos^2\alpha_0}\right)
=0.
\nonumber
\end{eqnarray}

We can now search for a perturbative instability, corresponding to an $E^2<0$ mode. The frequencies for
the $\delta\alpha_i$ modes are manifestly positive. The analysis of the $\delta\phi_1$, $\delta\alpha$
coupled system is not as simple. In what follows, we will restrict ourselves to small deformations $\tilde{\gamma}\ll 1$.
Obviously the positive energy modes can't become unstable for small $\tilde{\gamma}$, so that we focus on the zero modes.
The zero modes of the undeformed problem have $l=0$, so that we now focus on $l=0$. The $l=0$ modes satisfy

$$ \partial^2_0\delta\alpha
+A_1\delta\alpha +A_2\partial_0\delta\phi_1=0,$$

$$ \partial^2_0\delta\phi_1
-{A_2\over\cos^2\alpha_0}\partial_0\delta\alpha=0,$$

\noindent
In the undeformed case, where $A_1$ is zero, there are two zero modes corresponding to constant shifts in $\phi_1$
and $\alpha$. In the deformed case, $A_1<0$ so that although there is still a zero mode associated with constant
shifts of $\phi_1$, the zero mode associated with constant shifts of $\alpha$ is lifted\footnote{We thank the 
anonymous referee for comments which improved this presentation.}.

Even though the giant is perturbatively stable, it may still be unstable due to tunneling effects. To investigate this 
possibility, we look for bounce solutions of the Euclidean equations of motion. In the underformed case, instantons linking
the point graviton and sphere giants are known (see, for example, \cite{lee}). These solutions are obtained by allowing 
$\alpha$ (which determines the radius of the giant) to depend on time. Allowing both $\alpha$ and $\phi_1$ to depend on
time, after integrating over the spatial worldvolume coordinates, we find the Lagrangian ($r$ is the radius of the giant)

$$L=-m\sqrt{1-a\dot{\phi}^2_1-\dot{\alpha}^2}+b\dot{\phi}_1,$$

\noindent
where

$$m = N{r^3\over R^4} ,\qquad
a=R^2-r^2,\qquad r= R\sin\alpha_0$$

\begin{eqnarray}
b&=&4N\left[
{\tilde{\gamma}-\sqrt{4+\tilde{\gamma}^2}\over 4\tilde{\gamma}^2\sqrt{4+\tilde{\gamma}^2}}
\log\left(
{{2{r^2\over R^2}+{\sqrt{4+\tilde{\gamma}^2}\over\tilde{\gamma}}-1}\over
{{\sqrt{4+\tilde{\gamma}^2}\over\tilde{\gamma}}-1}}\right)-
{{\tilde{\gamma}+\sqrt{4+\tilde{\gamma}^2}\over 4\tilde{\gamma}^2\sqrt{4+\tilde{\gamma}^2}}}
\log\left(
{{-2{r^2\over R^2}+{\sqrt{4+\tilde{\gamma}^2}\over\tilde{\gamma}}+1}\over
{{\sqrt{4+\tilde{\gamma}^2}\over\tilde{\gamma}}+1}}\right)\right]
\nonumber\cr
&-&N{\tilde{\gamma}^2{r ^6 (R^2 -r^2)\over R^8 (1+\tilde{\gamma}^2{r^2\over R^2}(1-{r^2\over R^2}))}.}
\nonumber
\end{eqnarray}

\noindent
The canonical momenta are

$${\cal M}={\partial L\over\partial\dot{\phi}_1}={ma\dot{\phi}_1\over \sqrt{1-a\dot{\phi}^2_1-\dot{\alpha}^2}}+b,$$

$${\cal P}_\alpha={\partial L\over\partial\dot{\alpha}}={m\dot{\alpha}\over \sqrt{1-a\dot{\phi}^2_1-\dot{\alpha}^2}}.$$

\noindent
The Hamiltonian is obtained, as usual, by performing a Legendre transformation. In what follows, we treat the momentum
${\cal M}$ as a constant and make the Euclidean continuations ${\cal P}_\alpha^2\to -{\cal P}_\alpha^2$ and $H\to -H$
to obtain

$$H=-\sqrt{m^2 +{({\cal M}-b)^2\over a}-{\cal P}^2_\alpha}.$$

\noindent
The Euclidean equations of motion are now

$$\dot{\alpha}={\partial H\over \partial {\cal P}_\alpha},\qquad
\dot{\cal P}_\alpha=-{\partial H\over \partial \alpha}.$$

\noindent
These equations of motion are solved by ${\cal P}_\alpha=0$ and $\alpha=\alpha_0$ a constant with $\sin\alpha_0$ the
radius of the unstable giant. We have looked for numerical solutions to these equations by starting with
${\cal P}_\alpha=0$ and $\alpha=\alpha_0-\epsilon$ with $\epsilon\ll\alpha_0$. We find solutions as shown in
figure 4 below.

{\vskip -0.6cm}
\begin{center}
\begin{figure}[h]{\psfig{file=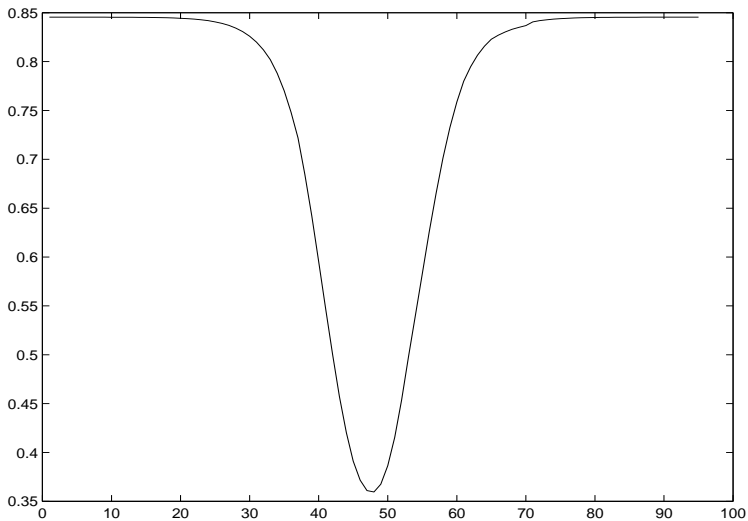,width=10cm,height=7cm}
 \caption{In the above plot $\alpha$ is shown as a function of $t$. The starting point 
is arbitrarily close to $\alpha_0$ where $\sin\alpha_0$ corresponds to the value of the
radius of the giant graviton for
$\tilde{\gamma}=0.4$, $N=10$ and ${\cal M}=7/N$.}}
\end{figure}
\end{center} 
{\vskip -0.6cm}

Our solutions are periodic with the period becoming arbitrarily long as we decrease the value of $\epsilon$. The value of $\alpha$
decreases to a minimum before returning to its initial value. These bounce solutions signal that our giant is unstable
due to tunneling effects\cite{bounce}.

\section{Open Strings}

The background studied in section 2 is conjectured\cite{F} to be dual to the field theory
with scalar potential 

$$ V =\Tr\sum_{n>m=1}^3 |e^{-i\pi\alpha_{mn}}\Phi_m\Phi_n-e^{i\pi\alpha_{mn}}\Phi_n\Phi_m|^2
+\Tr\sum_{n=1}^3\left[\Phi_n ,\bar{\Phi}_n\right]^2 ,$$

\noindent
where

$$ \alpha_{mn}=-\epsilon_{mni}\gamma_i . $$

\noindent
Below we will give a precise relation between the parameters $\gamma_i$ of the gauge theory and the parameters $\tilde{\gamma}_i$ of
the gravity background.
Our giant graviton solutions correspond to branes orbiting with angular momentum along the $\phi_1$ direction. 
The ${\cal R}$ charge of $\Phi_1$ corresponds to the angular momentum ${\cal M}$ of section 2. Thus, a giant
graviton with angular momentum ${\cal M}$ should be dual to an operator built out of ${\cal M}N$ $\Phi_1$
fields. From now on we use $Z$ to denote $\Phi_1$ and $X,Y$ to denote $\Phi_2,\Phi_3$.
To match what was done in the dual gravitational theory we set $\gamma_1=0$ and $\gamma_2=\gamma_3=\gamma$ so that

\begin{eqnarray}
V &=&\Tr\left[ |e^{i\pi\gamma}ZY-e^{-i\pi\gamma}YZ|^2+|e^{i\pi\gamma}XZ-e^{-i\pi\gamma}ZX|^2
+|YX-XY|^2\right. \nonumber\cr
&+& \left.\left[X ,\bar{X}\right]^2 +\left[Y ,\bar{Y}\right]^2+\left[Z ,\bar{Z}\right]^2\right].\nonumber
\end{eqnarray}

We would like to determine the spin chain of this deformed ${\cal N}=4$ super Yang-Mills theory
relevant for the dual description of open strings 
attached to giants. The spin chain for the deformed 
${\cal N}=4$ super Yang-Mills theory was found 
in \cite{rcb}; describing the open strings amounts to determining
what boundary conditions must be imposed on this spin chain. In the undeformed theory with
gauge group $U(N)$, operators dual to sphere giants 
are given by Schur polynomials of the totally antisymmetric
representations\cite{CJR}, which are labeled by 
Young diagrams with a single column. The cut off on the number of rows 
of the Young diagram perfectly matches the 
cut off on angular momentum arising because the sphere giant fills the $S^5$
of the AdS$_5\times$S$^5$ geometry. 
For maximal giants, the Schur polynomials are determinant like operators. Attaching
a string to the maximal giant gives an operator of the form

$${\cal O}=\epsilon^{j_1\cdots j_N}_{i_1\cdots i_N}Z^{i_1}_{j_1}\cdots Z^{i_{N-1}}_{j_{N-1}}
(M_1 M_2\cdots M_n)^{i_N}_{j_N}.$$

\noindent
The open string is given by the product $(M_1 M_2\cdots M_n)^{i_N}_{j_N}$. 
The $M_i$ could in principle
be fermions, covariant derivatives of Higgs fields or Higgs 
fields themselves. To describe excitations
of the string involving only coordinates from the $S^5$, 
we would restrict the $M_i$ to be Higgs fields.
We will restrict ourselves even further and require 
that the $M_i$ are $Z$ or $Y$. A spin chain description
can then be constructed by identifying 
$(M_1 M_2\cdots M_n)^{i_N}_{j_N}$ with a spin chain that has $n$-sites.
If $M_i=Z$ the $i$th spin is spin up; if $M_i=Y$ the 
$i$th spin is spin down. It is not possible for $Z$'s to
hop off and onto the string attached to a maximal 
giant; as soon as $M_1=Z$ or $M_n=Z$ the operator factorizes into
a closed string plus a maximal giant graviton. This implies 
the boundary constraint $M_1\ne Z\ne M_n$. However,
for non-maximal giants, $Z$'s can hop between the 
graviton and the open string. In this case, the number of sites
in the spin chain is dynamical. If however, one identifies the spaces 
between the $Y$'s as lattice sites and the
$Z$'s as bosons which occupy sites in this lattice, the 
number of sites is again conserved\cite{BCV}. For the undeformed
theory this leads to the Hamiltonian\cite{BCV}

$$ H=2\lambda\alpha^2 +2\lambda\sum_{l=1}^L\hat{a}_l^\dagger\hat{a}_l -\lambda 
\sum_{l=1}^{L-1}\left(\hat{a}_l^\dagger\hat{a}_{l+1}+
\hat{a}_l\hat{a}_{l+1}^\dagger\right)
+\lambda\alpha (\hat{a}_1+\hat{a}_1^\dagger)+\lambda\alpha (\hat{a}_L+\hat{a}_L^\dagger).$$

\noindent
The operators in the above Hamiltonian are Cuntz oscillators\cite{BCV}

$$a_ia_i^\dagger =I,\qquad a_i^\dagger a_i =I-|0\rangle\langle 0|.$$

\noindent
For a giant with angular momentum $p/R$, the parameter

$$\alpha =\sqrt{1-{p\over N}},$$

\noindent
measures how far from a maximal giant we are.

Due to the deformation, hopping is now accompanied by an extra phase. To see how this 
comes about, note that the deformation replaces

$$\left[ Z,Y\right]\to ZYe^{i\pi \gamma }-YZe^{-i\pi \gamma},$$
$$\left[ Z,Y\right]\left[ Z,Y\right]^\dagger \to
ZY\bar{Y}\bar{Z}+YZ\bar{Z}\bar{Y}-ZY\bar{Z}\bar{Y}e^{2\pi i\gamma}-YZ\bar{Y}\bar{Z}e^{-i2\pi\gamma}. $$

\noindent It is straight forward to see what 
interactions in the spin chain Hamiltonian these terms induce (the 
overbraces indicate Wick contractions) 

$$ \Tr (YZ\overbrace{\bar{Z}\overbrace{\bar{Y})\Tr(Y}Z}.....)\to \Tr (YZ...)\leftrightarrow a_l^\dagger a_l $$

$$\Tr (ZY\overbrace{\bar{Y}\overbrace{\bar{Z})\Tr (Z}Y}...)\to\Tr (ZY...)\leftrightarrow a_l^\dagger a_l $$

$$\Tr (ZY\overbrace{\bar{Z}\overbrace{\bar{Y}e^{i2\pi\gamma})\Tr (Y}Z}....)\to 
  e^{i2\pi\gamma}\Tr (ZY....)\leftrightarrow e^{i2\pi\gamma}a_l^\dagger a_{l+1} $$

$$\Tr (YZ\overbrace{\bar{Y}\overbrace{\bar{Z}e^{-i2\pi\gamma})\Tr (Z}Y}...)\to 
  e^{-i2\pi\gamma}\Tr (YZ...)\leftrightarrow e^{-i2\pi\gamma}a_l a_{l+1}^\dagger .$$

\noindent
To hop onto the spin chain, we are hopping from the ``zeroth site", which is the Schur polynomial/giant graviton, and onto the first 
site of the string. The term which does this has an $e^{-i2\pi\gamma}$ coefficient. Another way to hop onto the spin chain
is to hop from the $L+1$th site into the $L$th site. 
The term which does this has an $e^{i2\pi\gamma}$ coefficient. It is straight forward 
to argue for the phases when we hop off of the giant
graviton. From the above discussion we 
see that the deformation modifies this Hamiltonian to

\begin{eqnarray}
H &=& 2\lambda\alpha^2 +2\lambda\sum_{l=1}^L\hat{a}_l^\dagger\hat{a}_l -\lambda 
\sum_{l=1}^{L-1}\left(\hat{a}_l^\dagger\hat{a}_{l+1}e^{i2\pi\gamma}+
\hat{a}_l\hat{a}_{l+1}^\dagger e^{-i2\pi\gamma}\right)
\nonumber\\
&+& \lambda\alpha (\hat{a}_1e^{i2\pi\gamma}+\hat{a}_1^\dagger e^{-i2\pi\gamma})
+\lambda\alpha (\hat{a}_L e^{-i2\pi\gamma}+\hat{a}_L^\dagger e^{i2\pi\gamma}).
\nonumber
\end{eqnarray}

\noindent
In the above derivation of the deformed Hamiltonian we have considered only the
terms which look like $F$-terms. For this to be valid, it is necessary that the 
self energy, vector exchange and terms which look like $D$-terms, continue to cancel
as they did in the supersymmetric theory. It has been argued\cite{FRT2} that this is 
indeed the case, using the similarity between the $\beta$ deformation\cite{ls} and
non-commutative theories\cite{lb}.

The semiclassical limit, in which the action
derived from coherent states should provide a good approximation to the dynamics, is obtained by
taking

$$ L\sim\sqrt{N}\to\infty ,\qquad \lambda\to\infty,$$

\noindent
holding ${\lambda\over L^2}$, $L\gamma$ and $\alpha$ fixed. To obtain the low energy effective action, we 
will use the coherent states

$$ |z\rangle =\sqrt{1-|z|^2}\sum_{n=0}^\infty z^n|n\rangle ,$$

\noindent
with parameter 

$$ z_l=r_l e^{i\phi_l},$$

\noindent
for the $l$th site. The coherent state action is given as usual by

$$ S=\int dt\left(i\langle Z|{\partial\over\partial t}|Z\rangle -\langle Z|H|Z\rangle \right).$$

\noindent
In the above expression the coherent state $|Z\rangle$ is written as a product over all sites

$$|Z\rangle =\prod_{l}|z_l\rangle .$$

\noindent
As an illustration of the manipulations which follow, we describe the evaluation of the first term in the action.
It is straight forward to see that

$$ {\partial\over\partial t}|z_l\rangle 
=-{r_l\dot{r}_l\over\sqrt{1-r_l^2}}\sum_{n=0}^\infty r_l^ne^{in\phi_l} |n\rangle
+\sqrt{1-r_l^2}\sum_{n=0}^\infty n\left({\dot{r}_l\over r_l}+i\dot{\phi}_l\right)
r_l^ne^{in\phi_l} |n\rangle ,$$

$$\langle z_m|{\partial\over\partial t}|z_l\rangle 
=i{r_l^2\dot{\phi}_l\over 1-r_l^2 }\delta_{lm}.$$

\noindent
Thus,

$$\langle Z|{\partial\over\partial t}|Z\rangle 
=i\sum_{l=1}^L {r_l^2\dot{\phi}_l\over 1-r_l^2 }.$$

\noindent
In the large $L$ limit, to leading order in $L$ we have

$$\langle Z|{\partial\over\partial t}|Z\rangle 
=iL\int_{0}^1 {r(\sigma)^2\dot{\phi}(\sigma)\over 1-r(\sigma)^2 }d\sigma  .$$

\noindent
A straight forward computation along these lines gives

\begin{eqnarray}
S&=&-\int dt\left[L \int_0^1 {r^2\dot{\phi}\over 1-r^2 }d\sigma + 2\lambda\alpha^2
+{\lambda\over L}\int_0^1 \left( (r')^2 +r^2 (\phi '+2\pi\tilde{\gamma})^2 \right)d\sigma\right.
\nonumber\\
&+&\lambda \bar{z}(1) z(1) +\lambda\bar{z}(0)z(0) \left. +\lambda\alpha (z(0)+\bar{z}(0))
+\lambda\alpha (z(1)+\bar{z}(1))\right].\nonumber
\end{eqnarray}

\noindent
We identify

$$\tilde{\gamma}=L\gamma .$$

\noindent
Write this action in terms of $\gamma$ and rescale $\sigma\to{\sigma\over\pi}$. Clearly, the deformation
replaces

$$\phi'\to\phi '+2 L\gamma .$$
 
Lets now consider the description of the open strings using the dual sigma model. The undeformed
case has been studied in\cite{BCV}. The work \cite{BCV}
uses a coordinate system in which the brane is static, a gauge in which $p_{\phi_2}$ is
homogeneously distributed along the string, $p_{\phi_2}=2{\cal J}$ and $\tau =t$.
After taking a low energy limit, the string sigma model action is

$$ -\sqrt{\lambda_{YM}}\int dt\int_0^{2\pi} {d\sigma\over 2\pi}\left[
{r^2\dot{\phi}_1\over 1-r^2}+{\lambda_{YM}\over 8\pi^2{\cal J}^2}(r^{\prime 2}+r^2
\phi_1^{\prime 2})\right],$$

\noindent
in perfect agreement with the undeformed result from the field theory\cite{BCV},
after identifying $L={\cal J}$ and $\lambda_{YM}=8\pi^2\lambda $. 

The background studied in section 2 can be obtained by performing a sequence of TsT
transformations\cite{F}. A TsT transformation exploits a two torus,
with coordinates $(\phi_1,\phi_2)$ say, in the geometry. A TsT transformation begins
with a $T$-duality with respect to $\phi_1$, then a shift $\phi_2\to\phi_2+\gamma\phi_1$
and finally a second $T$-duality along $\phi_1$. In the $AdS_5\times S^5$ background there are three
natural tori  $(\phi_1,\phi_2)$ $(\phi_2,\phi_3)$ and $(\phi_3,\phi_1)$. This allows three
independent TsT transformations giving the three parameter deformation of section 2. See
\cite{F} for details. The TsT transformation has a particularly simple action on the string sigma
model, something which was exploited in\cite{F} to obtain the Lax pair for the bosonic part of the sigma model.
To obtain the sigma model for the deformed theory, we simply need to shift\cite{F}

$$\phi_i'\to \phi_i' -\epsilon_{ijk}\gamma_j p_k .$$

\noindent
For the above action, we only need to consider $\phi_1'$

$$\phi_1'\to \phi_1' -\epsilon_{1jk}\gamma_j p_k .$$

\noindent
Next, since we set $X=0$ we know that $p_3=0$. Thus,

$$\phi_1'\to \phi_1' -\epsilon_{1j2}\gamma_j p_2 = \phi_1' -\epsilon_{132}\gamma_3 p_2 =
\phi_1' + \gamma_3 p_2.$$

\noindent
Now, we have set $\gamma_3=\gamma_2=\gamma $ and in our gauge $p_2 = 2{\cal J}$, so that

$$\phi_1'\to \phi_1' + 2\gamma {\cal J} =\phi_1' + 2\gamma L .$$

\noindent
This is in complete agreement with the spin chain result. 

\section{Summary}

In this paper we have found giant graviton solutions in the deformed background
with $\tilde{\gamma}_1=0$ and $\tilde{\gamma}_2=\tilde{\gamma}_3\equiv\tilde{\gamma}$.
These giants have an energy which is greater than the energy of a point graviton. We have
also considered the spectrum of small fluctuations about these giants. The spectrum depends
on the radius of the giant in contrast to the undeformed case where the spectrum is independent 
of the size of the giant\cite{djm}. For small deformations, we have argued that the giant graviton 
is perturbatively stable. The Euclidean equations of motion admit a bounce solution indicating
that the giant graviton will be unstable due to tunneling effects.
We have also considered the semiclassical dynamics of open strings attached
to these giants. We find that there is perfect quantitative agreement between the gauge
theory and the string theory. Indeed, the deformation in the gauge theory exactly
reproduces the TsT transformation relating the deformed and undeformed sigma models!

The comparison in this paper provides further quantitative agreement following from
AdS/CFT duality in a non-supersymmetric case. Further, the fact that the giant graviton is
unstable makes the quantitative agreement even more interesting.

There are a number of directions in which the present work can be extended. It would
be interesting to look for giant gravitons in the general three parameter deformed background.
One could also consider giants which have expanded into the AdS$_5$ space; the giant will be the same 
as the solution presented in\cite{AdS}; the deformation should however modify the small fluctuation
spectrum\cite{djm}. Further, the open string fluctuations we have considered are certainly
not the most general fluctuations that can be considered. It would be interesting to extend
our results to see if the agreement we have found continues to hold for more general
open string configurations.

$$ $$

\noindent
{\it Acknowledgements:} We would like to thank Rajsekhar Bhattacharyya and Jeff Murugan
for pleasant discussions. This work is supported by NRF grant number Gun 2047219.

\end{document}